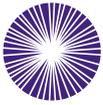
AMERICAN ACADEMY OF OPHTHALMOLOGY®

# Perfused and Nonperfused Microaneurysms Identified and Characterized by Structural and Angiographic OCT

*Min Gao, MS,*[1] *Tristan T. Hormel, PhD,*[1] *Yukun Guo, MS,*[1,2] *Kotaro Tsuboi, MD,*[1] *Christina J. Flaxel, MD,*[1] *David Huang, MD, PhD,*[1,2] *Thomas S. Hwang, MD,*[1] *Yali Jia, PhD*[1,2]

***Purpose:*** Microaneurysms (MAs) have distinct, oval-shaped, hyperreflective walls on structural OCT, and inconsistent flow signal in the lumen with OCT angiography (OCTA). Their relationship to regional macular edema in diabetic retinopathy (DR) has not been quantitatively explored.

***Design:*** Retrospective, cross-sectional study.

***Participants:*** A total of 99 participants, including 23 with mild, nonproliferative DR (NPDR), 25 with moderate NPDR, 34 with severe NPDR, and 17 with proliferative DR.

***Methods:*** We obtained 3 × 3-mm scans with a commercial device (Solix, Visionix/Optovue) in 99 patients with DR. Trained graders manually identified MAs and their location relative to the anatomic layers from cross-sectional OCT. Microaneurysms were first classified as perfused if flow signal was present in the OCTA channel. Then, perfused MAs were further classified into fully and partially perfused MAs based on the flow characteristics in *en face* OCTA. The presence of retinal fluid based on OCT near MAs was compared between perfused and nonperfused types. We also compared OCT-based MA detection to fundus photography (FP)- and fluorescein angiography (FA)-based detection.

***Main Outcome Measures:*** OCT-identified MAs can be classified according to colocalized OCTA flow signal into fully perfused, partially perfused, and nonperfused types. Fully perfused MAs may be more likely to be associated with diabetic macular edema (DME) than those without flow.

***Results:*** We identified 308 MAs (166 fully perfused, 88 partially perfused, 54 nonperfused) in 42 eyes using OCT and OCTA. Nearly half of the MAs identified in this study straddle the inner nuclear layer and outer plexiform layer. Compared with partially perfused and nonperfused MAs, fully perfused MAs were more likely to be associated with local retinal fluid. The associated fluid volumes were larger with fully perfused MAs compared with other types. OCT/OCTA detected all MAs found on FP. Although not all MAs seen with FA were identified with OCT, some MAs seen with OCT were not visible with FA or FP.

***Conclusions:*** OCT-identified MAs with colocalized flow on OCTA are more likely to be associated with DME than those without flow.

***Financial Disclosure(s):*** Proprietary or commercial disclosure may be found in the Footnotes and Disclosures at the end of this article. *Ophthalmology Retina* 2023;■:1–8 © 2023 by the American Academy of Ophthalmology. This is an open access article under the CC BY-NC-ND license (http://creativecommons.org/licenses/by-nc-nd/4.0/).

*Supplemental material available at* www.opthalmologyretina.org.

Diabetic retinopathy (DR) is a leading cause of vision loss.[1,2] Microaneurysms (MAs) are a key feature of DR[3] and play an important role in the development of diabetic macular edema (DME), the most frequent cause of vision loss in DR.[4] The ETDRS and Diabetic Retinopathy Clinical Research Network used a focal laser protocol for DME that targeted MAs and demonstrated the strategy to be effective.[5,6]

Microaneurysms can be visualized using several imaging modalities, including fundus photography (FP),[7–11] fluorescein angiography (FA),[12–14] OCT,[13,15] and OCT angiography (OCTA),[16,17] each with unique advantages and disadvantages. The pinpoint hyperfluorescence with associated leakage on an FA is a classic description of an MA, and it has been demonstrated to be more sensitive than FP for detection of MAs.[18] However, with the requirement of intravenous dye injection, it is not routinely used as a screening modality. Structural OCT can consistently detect MAs as oval structures with hyperreflective walls, consistent with histopathologic findings.[13,15,19] OCT can also distinguish MAs from "dot hemorrhages," which are small, intraretinal hemorrhages that are ophthalmoscopically indistinguishable from MAs.[20] On OCTA, instead of appearing as a pinpoint of flow, MAs appear as focal, saccular, fusiform, or pedunculated lesions connected to capillaries, demonstrating detailed morphology analogous to histopathologic descriptions.[17] However, not all MAs identified with FA can be seen on OCTA, although averaging multiple scans can increase the yield.[17,21] OCT angiography also lacks dynamic information, such as leakage, used to guide laser photocoagulation for DME.[22] Despite the wide range of imaging modalities available, and the known relationship to DME,[23,24] no study to our knowledge has established a local relationship between MA characteristics and DME quantitatively.

In this study, we combined OCTA data, with its inherently coregistered structural OCT data, to explore whether flow characteristics are related to MA pathology. Specifically, we asked whether MAs identified are more likely to be associated with edema based on flow characteristics. We also investigated the location of the MAs relative to the retinal layers, and explored the consistency of MA identification between OCT/OCTA and other modalities (FP and FA).

## Methods

### Data Acquisition

The institutional review board of Oregon Health & Science University approved the study. Informed consent was obtained from all participants, and the study adhered to the Declaration of Helsinki. The severity of DR was determined by clinical examination (T.T.H.) according to the ETDRS scale.[25] Eight 3 × 3-mm OCT and OCTA scans with a 400 × 400 sampling density from 1 and only 1 eye of each participant were obtained using a commercial 120-kHz spectral domain-OCT system (Solix; Visionix/Optovue, Inc). The split-spectrum amplitude-decorrelation angiography algorithm implemented in this instrument was used to generate the OCTA data.[26] Four pairs of the orthogonal scans (x-fast and y-fast) (Fig S1A, B, available at www.ophthalmologyretina.org) were registered to generate 4 motion-free volumes (Fig S1C). Then these 4 volumes were registered and merged together to obtain a high-definition volume (Fig S1D).[27] A projection-resolved OCTA algorithm was applied to suppress projection artifacts through the entire volume.[28] The internal limiting membrane, nerve fiber layer, ganglion cell/inner plexiform layer, inner nuclear layer (INL), outer plexiform layer (OPL), and the outer nuclear layer were segmented by a guided, bidirectional graph search algorithm.[29] Fifty-degree color fundus images (2392 × 2048 pixels) were captured using a fundus camera (FF450, Carl Zeiss Meditec AG), and 200-degree FA (4000 × 4000 pixels) was obtained using a commercial system (P200DTx, Optos plc, Dunfermline).

### Identification of MAs with OCT

Two masked graders (M.G. and Y.G.) manually identified and delineated MAs that appeared as well-demarcated round or oval lesions with hyperreflective walls visible with cross-sectional OCT (Fig S2, A1, A2, available at www.ophthalmologyretina.org).[13,15] Every B-scan was examined for each scan volume yielding voxel-resolution MA volumes when combined.

To investigate the MA distribution within the macula, we calculated the volume and number of MAs in each retinal layer (nerve fiber layer, ganglion cell/inner plexiform layer, INL, OPL, and outer nuclear layer). The MA volume proportion in each retinal layer was defined as the ratio of MA volume in each layer to the total MA volume. If the MA volume proportion in any given MA within a layer was > 10%, then that MA was counted as being located within the layer. In this way, many MAs were present in multiple layers.

### Classification of MAs from OCTA Overlaid on OCT

To create a clean slab that demonstrates MAs while suppressing noise, we defined the anterior and posterior boundaries of the slab based on the locations of most anterior and most posterior structural MA voxels. The mean structural OCT and maximum OCTA signal from the same slab were projected to generate 2-dimensional *en face* images (Fig S2, B1−C2). If flow signal was present within an MA in cross-sectional OCTA, then we segmented the *en face* OCTA flow signal coregistered to the structural OCT scan from the same slab.

First, we classified MAs as perfused versus nonperfused, depending on the presence of flow signal from cross-sectional OCTA within the lumen of structural MAs. Then we further classified perfused MAs as fully perfused if the characteristic vascular outpouching can be observed on *en face* OCTA image, and partially perfused when blood flow is present on both cross-sectional and *en face* OCTA images, but the flow signal is weak and less discernable. We used a Kolmogorov−Smirnov test to determine whether MA volume in the OCT channel was normally distributed then a nonparametric Kruskal−Wallis test to determine whether the structural volume of 3 types of MAs (fully perfused, partially perfused, and nonperfused MAs) have significant differences. Finally, a post hoc test with Bonferroni correction was performed to determine significant differences between pairs of comparison (nonperfused vs. partially perfused; nonperfused vs. fully perfused; partially perfused vs. fully perfused MAs).

### Investigation of Relations of MAs with Retinal Fluid

We also investigated the local relationship between the types of MAs and the intraretinal fluid volume within 225 μm (about 2 times the mean diameter of MAs in *en face* structural OCT) from the manually delineated edge of the MAs. A convolutional neural network-based algorithm we previously developed automatically quantified the intraretinal cystoid fluid volume (Fig S2, D1, D2).[30]

We applied a chi-square test to compare the presence probabilities of fluid near 3 types of MAs. We used a Kolmogorov−Smirnov test to determine whether the fluid volume near MAs was normally distributed, and a nonparametric Kruskal−Wallis test to determine whether the fluid volume near these 3 types of MAs had significant differences. Finally, a post hoc test with Bonferroni correction was performed to test for significant differences between pairwise comparisons between the 3 types of MAs.

### Comparison of MAs between Different Imaging Modalities

A portion of the eyes also underwent FP and FA imaging. Two graders (M.G. and K.T.) identified MAs as dark-red, isolated dots on FP and hyperfluorescent dots on consecutively captured 5 to 10 FA images over a duration of approximately 10 minutes. Frames with low quality were excluded, and the graders would reach a consensus through discussion. Then we compared the number of MAs detected by FP, FA, and OCT/OCTA and explored reasons for inconsistency between modalities. We analyzed the capsular MAs with a bright wall in cross-sectional OCT, the flow perfusion status of MAs in OCTA, and the hyperfluorescent dots in FA to characterize features that enabled detection with both modalities. We also investigated how these features relate to MAs visible with FP. We created an MA typology based on our observations.

## Results

### Patient Characteristics

This study included 99 eyes (23 with mild nonproliferative DR [NPDR], 25 with moderate NPDR, 34 with severe NPDR, and 17 with proliferative DR). Microaneurysms were identified in 42 eyes (13 [56.5%] with mild NPDR, 7 [28.0%] with moderate NPDR, 13 [38.2%] with severe NPDR, and 9 [52.9%] with proliferative DR). Of the 42 eyes, 23 had fundus photographs and 14 had fluorescein angiograms. These eyes were used for comparison to other modalities.

### Distribution of MAs across Retinal Layers

We identified 308 MAs in 42 eyes based on highly reflective circular or elliptical walls seen on cross-sectional structural OCT. We found that MAs were present in multiple retinal layers, including the nerve fiber layer, ganglion cell/inner plexiform layer, INL, OPL, and outer nuclear layer (Fig S3, available at www.ophthalmologyretina.org). By measuring the number and volume of MAs in each retinal layer, we found that MAs were most likely to be in INL and OPL, with the highest volume also in those layers (Fig S3A, B). Most MAs spanned multiple layers, with 46% of all MAs spanning INL and OPL (Fig S3C, D).

### Characteristics of MAs in OCT and OCTA

Microaneurysms identified on the cross-sectional OCT (Fig 4A) corresponded to very bright round or irregular spots on *en face* OCT images (Fig 4B). The perfusion status of MAs can be determined by the coregistered, cross-sectional OCTA (Fig 4C). The flow perfused the lumen inside the MAs walls; however, some MAs had no discernable flow in OCTA channel (Fig 4D).

### Perfused and Nonperfused MAs

Of 308 MAs, 254 MAs were classified as perfused based on the flow signal seen with the cross-sectional OCTA (Fig 5). These perfused MAs were further identified as 166 fully perfused MAs (Fig 5, D1, E1) and 88 partially perfused MAs (Fig 5, D2, E2) based on the flow morphology of MAs on *en face* OCTA.

In terms of mean volume, the fully perfused MAs were significantly larger than partially perfused (volume [$\times 10^5$ μm$^3$]: 4.5 ± 0.4 vs. 2.8 ± 0.2, $P < 0.001$) and nonperfused MAs (volume [$\times 10^5$ μm$^3$]: 4.5 ± 0.4 vs. 3.0 ± 0.5, $P = 0.001$) in the OCT channel. There was no significant difference in the mean (± standard deviation) volume between partially perfused MAs and nonperfused MAs (volume [$\times 10^5$ μm$^3$]: 2.8 ± 0.2 vs. 3.0 ± 0.5, $P = 1.000$).

### Correlation between Perfused MAs and Other DR Complications

The presence probabilities of fluid near the fully perfused, partially perfused, and nonperfused MAs were 80.1% (133/166), 63.6% (56/88), and 63.0% (34/54), respectively. A significant association was found between the types of MA and the presence of fluid (chi-square test, $P = 0.005$; Fig 6). Retinal fluid is much more likely to be present near fully perfused MAs than partially perfused and nonperfused MAs. A post hoc test with Bonferroni correction was applied to the pairwise comparison of fluid near these 3 types of MAs, which showed that the mean (± standard deviation) retinal fluid volume near the fully perfused MAs was significantly larger compared with the partially perfused MAs (volume [$\times 10^6$ μm$^3$]: 5.0 ± 0.5 vs. 3.1 ± 0.6, $P = 0.021$) and nonperfused MAs (volume [$\times 10^6$ μm$^3$]: 5.0 ± 0.5 vs. 2.1 ± 0.4, $P = 0.042$); however, there was no significant difference between the partially perfused MAs and nonperfused MAs (volume [$\times 10^6$ μm$^3$]: 3.1 ± 0.6 vs. 2.1 ± 0.4, $P = 1.000$).

### Comparison of the Detection of MAs by OCT/OCTA, FP, and FA

We explored the agreement between OCT, OCTA, FP, and FA by counting the number of MAs seen on each modality within the same 3 × 3-mm field of view. In the 23 eyes that also had FP images, FP (Fig S7, D1, D2, available at www.ophthalmologyretina.org) captured 62 (44.3%) of 140 perfused MAs (Fig S7, C1, C2) and 13 of 35 (37.1%) nonperfused MAs that appeared in OCT (Fig S7, A1−B2). No definite MAs were seen on FP that OCT or OCTA did not identify. In the 14 eyes that also had FA images, FA (Fig S7, E1, E2) demonstrated 56 (64.3%) of 87 perfused MAs and 15 (68.2%) of 22 nonperfused MAs identified with OCT. OCT detected 71 (41.0%) of 173 MAs that appeared on FA, 56 of which were perfused MAs and 15 nonperfused MAs.

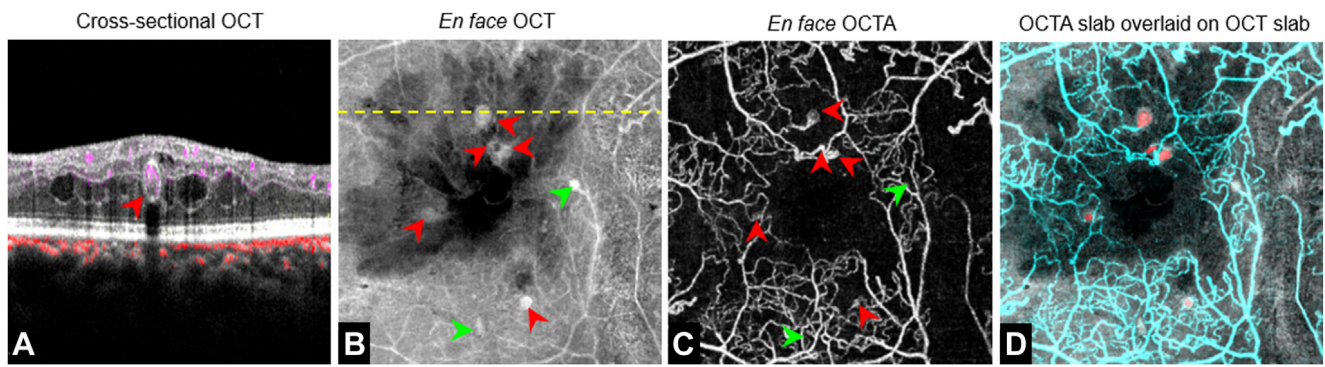

**Figure 4.** Characterization of microaneurysms (MAs) in an eye with macular edema. **A,** An MA presents as an oval structure with a strongly reflective wall (red arrow) and flow in cross-sectional structural OCT/OCT angiography (OCTA). **B,** The lesions appear as bright round spots in *en face* OCT. The dashed yellow line indicates the location of the cross-sectional image. **C,** The *en face* OCTA shows flow signal in some MAs (red arrows), whereas some MAs don't contain flow signal (green arrows). **D,** There is very discernable flow (red) in some MAs shown by overlaid *en face* OCT (gray) and OCTA (**C**).

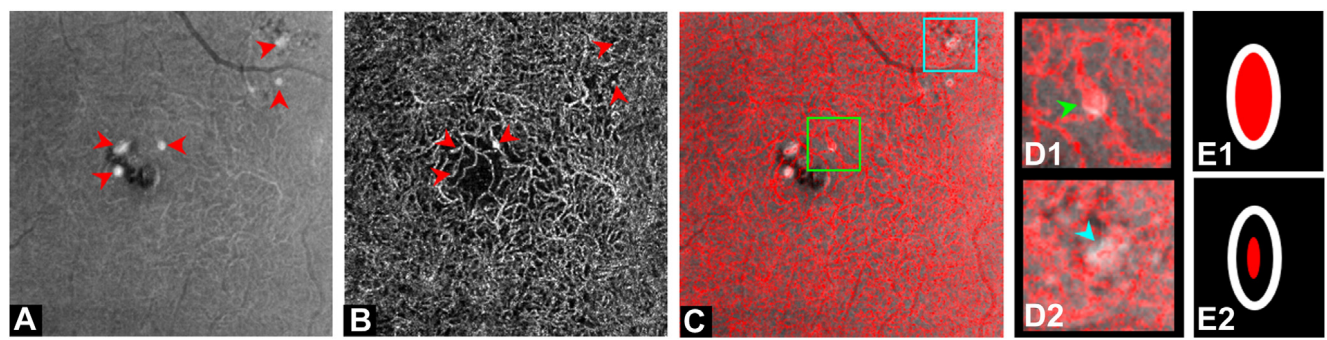

**Figure 5.** Flow perfusion status in microaneurysms (MAs) from an eye with mild nonproliferative diabetic retinopathy. **A,** Microaneurysms (red arrows) in an *en face* OCT image. **B,** Microaneurysms (red arrows) in an *en face* OCTA angiogram. (C) *En face* OCT angiography (OCTA) (red) overlaid on *en face* OCT (gray). **D1, 2,** Magnified views from the green and cyan boxes in **(C)** of overlaid *en face* images. A fully perfused MA (green arrow) shows a vascular protrusion on *en face* OCTA, and a partially perfused MA (cyan arrow) does not appear as an outpouching of the vessel on *en face* OCTA. **E1, 2,** Illustration of a fully perfused MA and a partially perfused MA in the lumen in cross-sectional OCT and OCTA. The white oval edge represents the lumen wall. The red ellipse refers to the flow perfusion status inside MAs.

Comparing the detection of MAs with OCT, OCTA, and FA, we found 6 specific scenarios. First, some MAs were seen in every modality, with distinct hyperreflective walls on OCT, flow signal on OCTA, and focal hyperfluorescence on FA (Fig 8, row 1). There were also MAs that were seen on OCT and OCTA but without corresponding hyperfluorescence on FA (Fig 8, row 2). These could represent MAs that are occluded from circulation but have enough lumen to allow adequate movement of red blood cells that OCTA could detect. Another group of MAs could be detected with OCT and FA but not with OCTA (Fig 8, row 3). These may be MAs that are partially occluded or have lumen that is too small to allow for movement of red blood cells that can be detected by OCTA, but still connected to circulation to allow entry and accumulation of fluorescein dye. There were also MAs that were only seen by OCT with characteristic hyperreflective walls but not with OCTA or FA (Fig 8, row 4). These MAs are likely completely occluded or clotted, not allowing flow or movement. A few MAs were seen on FA but not on OCT (Fig 8, rows 5 and 6). These are likely small lesions with walls not thick enough to be characterized by hyperreflective walls on OCT. Some of these had flow signal on OCTA. No single modality was able to capture all lesions that have been clinically and histopathologically described as MAs.

## Discussion

In this work, we used OCTA and its coregistered structural OCT to characterize MAs in DR, and quantitatively examined their relationship to retinal edema. Other groups have also characterized MAs by the observation of OCT and

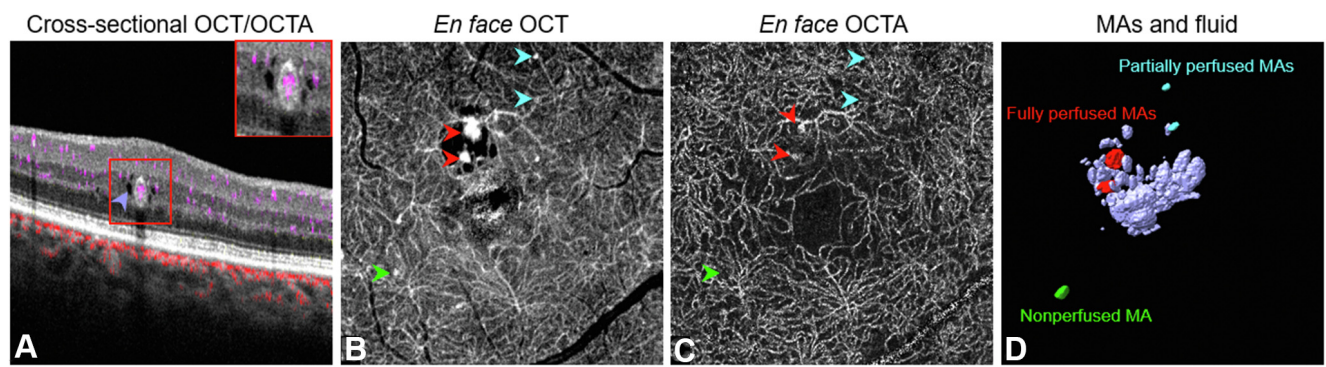

**Figure 6.** The relationship between microaneurysms (MAs) and nearby fluid. **A,** Fluid (purple arrow) is presented near a fully perfused MA (red box) on the cross-sectional OCT/OCT angiography (OCTA). **B,** Microaneurysms identified in an *en face* OCT image from an eye with mild nonproliferative diabetic retinopathy. **C,** Two fully perfused (red arrows) and 2 partially perfused MAs (cyan arrows) are visible on *en face* OCTA angiogram. A nonperfused MA (green arrow) does not show flow. **D,** Retinal fluid (purple) overlapped with 3 types of MAs. The retinal fluid volume near the fully perfused MAs was significantly larger than that near the partially perfused and nonperfused MAs.

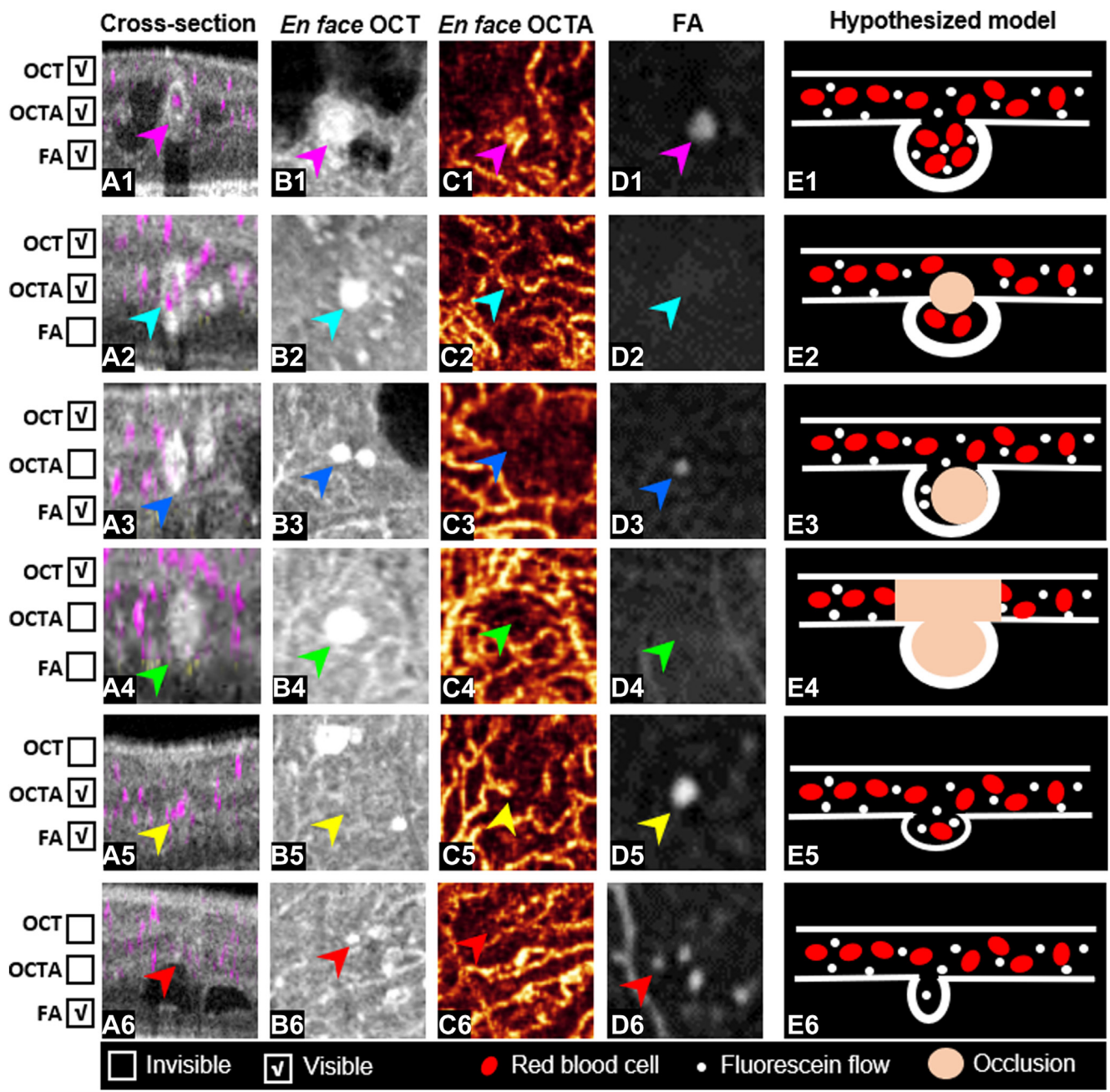

**Figure 8.** Comparison of microaneurysm (MA) detection with cross-sectional OCT angiography (OCTA) (column 1), *en face* OCT (column 2), *en face* OCTA (column 3), and fluorescein angiography (FA) (column 4). Each row demonstrates different situations where the detection with these modalities vary and the potential explanation for the discordance (column 5): (Row 1) The flow within the hyperreflective lumen is detected by OCTA and FA; (Row 2) The flow into MA is occluded, precluding entry of fluorescein but there is detectable movement within the lumen; (Row 3) A partially occluded MA allows fluorescein into lumen, but there is inadequate lumen to allow movement that is detected by OCTA; (Row 4) The lumen is completed occluded and there is no entry of fluorescein or movement of cells possible; (Row 5) The wall of MA is too thin to be seen on OCT, but the lumen is adequate to allow movement of cells and entry of fluorescein; (Row 6) Similar to row 5, but the lumen is too small to allow movement of cells.

OCTA separately. Fukuda et al[21] averaged multiple OCTA volumes to classify MAs based on OCTA morphology and found that fusiform MAs were more associated with leakage on FA than others. This approach, however, only identifies MAs seen on OCTA, which, in our study, identified fewer MAs compared with other modalities and requires a subjective evaluation of the shape. Although our approach also required manual identification and segmentation of the MAs, we classified lesions based on the presence or absence of colocalized flow signal, which may be more objective than morphological classification. Parravano et al[31,32] found the hyperreflective MAs on cross-sectional

OCT were associated with MA detection on OCTA, and they also found that the visibility and the location of MAs on OCTA were strongly associated with extracellular fluid accumulation at 1 year. However, MA flow was studied in low-definition 6 × 6-mm single OCTA scans, and OCTA wasn't registered on OCT, which may cause MAs with a weak flow to be missed. Kaizu et al[17] investigated MA detection using multiple en face OCTA image averaging. They concluded that multiple image averaging is useful for increasing the MA detection capability of OCTA. We fully agree on this point; a high-definition scan generated by 8 volumes was investigated on each eye. Khatri et al[33] and Chen et al[34] also analyzed flow in MAs using OCTA and concluded that the MAs with flow presence were associated with cystoid spaces. Before the advent of OCTA, Soliman et al[35] described the pattern of MAs on structural OCT and validated them with FA on the same eyes. However, they did not offer a quantitative analysis of the relationship between MAs and retinal edema; also, they cannot reveal the flow status within MAs due to the lack of OCTA.

In this study, we localized the MAs with respect to the retinal layers on structural OCT and found that most MAs are in the INL and OPL, with a majority spanning multiple layers. This is consistent with a previous immunohistologic analysis that found MAs in DR originated predominantly from the INL.[36] We did not try to classify the MAs as belonging to the superficial or deep vascular complex. Unlike the normal capillary plexuses that lie in a relatively flat configuration with separation between the layers and a few connecting vessels (features that lend themselves to segmentation by retinal layers), MAs span multiple layers and defy equivalent segmentation. This explains why previous studies that used the superficial versus deep vascular complex segmentation to demonstrate MAs frequently found that MAs were inconsistently seen in 1 of the layers or both layers.

Our findings suggest that FA, long considered the gold standard for the detection of MAs, fails to detect some MAs visible with OCT and OCTA. When we compared the detection of MAs with multiple modalities (FA, OCT, and OCTA), there was significant variability, with no one modality emerging as clearly superior. Further study can elucidate the clinical significance of this variability.

Our study included several novel characterizations. First, the combination of OCT and OCTA allowed us to study the flow status of MAs detected by OCT. This helped to elucidate the relation between morphology and flow in vivo, which cannot be done by histology using the fixation of retinal tissue and blood cells. Second, we provided clear guidance on grading MAs imaged with structural OCT, i.e., which layers should be given extra consideration and which features are relevant. We also divided MAs detected by OCT into 3 groups based on a distinguishable OCTA signal in en face images, which is a more practical way to stratify their perfusion status. Third, we compared OCT/OCTA detected MAs to FA and FP and interpreted the incongruity between them, which can help explain disparities between the MAs visualized by the different modalities. Finally, we investigated the relationship between the local fluid and the types of MAs; perfused flow is associated with retinal fluid accumulation. These findings indicate that OCT/OCTA evaluation of MAs may help plan the treatment of DME. Overall, to the best of our knowledge, this is the pioneering study to quantitatively demonstrate the relationship between an imaging characteristic of MAs (perfusion status) and retinal edema in the surrounding area. This relationship not only confirms the conventional wisdom about the relationship between DME and MAs, but also suggests that this particular imaging characteristic may predict the development of DME.

Limitations of this study include its cross-sectional and retrospective nature, and the relatively small number of patients examined. Also, the manual nature of identification and segmentation makes the findings in this study impractical for clinical practice. Deep learning techniques could be applied in the future to automatically segment these lesions so that the perfused MAs could be presented to the clinicians without the need to examine individual cross-sectional OCT scans to compare to en face OCTA. Although the study did find a significant relationship between fully perfused MAs and macular edema, it did not examine whether output such as this from OCT and OCTA could be used to guide laser photocoagulation for DME in place of FA. A prospective study that compares the use of OCTA versus FA would be required to validate such a use. Finally, a prospective study is necessary to evaluate the value of perfused MAs as identified by OCTA in predicting the development of DME. Although this study suggests that there is a relationship between DME and perfused MAs, whether this relationship is clinically meaningful remains to be seen. In addition, further studies are required to determine whether there are differences in flow specific to certain OCTA processing algorithms, as various OCTA devices may have distinct processing algorithms.

In conclusion, a coupled structural and angiographic OCT can characterize MAs based on colocalized flow. The MAs with flow are more likely to be associated with local edema. Further study is needed to evaluate whether this biomarker could be used to guide treatment or predict the development of DME.

## Footnotes and Disclosures

[1] Casey Eye Institute, Oregon Health & Science University, Portland, Oregon.

[2] Department of Biomedical Engineering, Oregon Health & Science University, Portland, Oregon.

Disclosure(s):

The author(s) have made the following disclosure(s): Y.G: Patent — Visionix/Optovue Inc., Genentech Inc.

K.T.: Honoraria — Bayer, Santen, Chugai, Alcon, Senju

D.H.: Financial interest — Visonix/Optovue Inc.; Patent — Visionix/Optovue Inc.; Recipient — Visionix/Optovue Inc; Consultant — Boeringer Ingelheim Inc.

Y.J.: Financial interest — Visonix/Optovue Inc.; Patent — Visionix/Optovue, Inc., Optos. Inc., Genentech Inc.

The other authors have no proprietary or commercial interest in any materials discussed in this article.

This work was supported by grant National Institutes of Health (R01 EY035410, R01 EY027833, R01 EY024544, R01 EY031394, P30 EY010572, T32 EY023211, UL1TR002369); the Malcolm M. Marquis, MD, Endowed Fund for Innovation; Unrestricted Departmental Funding Grant and Dr. H. James and Carole Free Catalyst Award from Research to Prevent Blindness (New York, NY); Edward N. & Della L. Thome Memorial Foundation Award; and Bright Focus Foundation (G2020168, M20230081).

Meeting presentation: This research was presented at the Association for Research in Vision and Ophthalmology (ARVO), May 1–4, 2022 in Denver, CO (Members-in-Training Outstanding Poster).

HUMAN SUBJECTS: Human subjects were included in this study. The institutional review board of Oregon Health & Science University approved the study. Informed consent was obtained from all participants and the study adhered to the tenets of the Declaration of Helsinki.

No animal subjects were used in this study.

Author Contributions:

Conception and design: Gao, Jia

Data collection: Gao, Guo, Tsuboi, Hwang

Analysis and interpretation: Gao, Hormel, Guo

Obtained funding: Jia

Overall responsibility: Gao, Hormel, Flaxel, Huang, Hwang, Jia

Abbreviations and Acronyms:

**DME** = diabetic macular edema; **DR** = diabetic retinopathy; **FA** = fluorescein angiography; **FP** = fundus photography; **INL** = inner nuclear layer; **MA** = microaneurysm; **NPDR** = nonproliferative diabetic retinopathy; **OCTA** = OCT angiography; **OPL** = outer plexiform layer.

Keywords:

OCT, OCT angiography, Perfused and nonperfused microaneurysms.

Correspondence:

Yali Jia, PhD, Casey Eye Institute, Oregon Health & Science University, 515 SW Campus Dr, Portland, OR 97239. E-mail: jiaya@ohsu.edu.

# References

1. Batchelder T, Barricks M. The Wisconsin epidemiologic study of diabetic retinopathy. *Arch Ophthalmol*. 1995;113:520–526.
2. Stitt AW, Curtis TM, Chen M, et al. The progress in understanding and treatment of diabetic retinopathy. *Prog Retin Eye Res*. 2016;51:156–186.
3. Wong TY, Klein R, Sharrett AR, et al. Retinal microvascular abnormalities and cognitive impairment in middle-aged persons: the Atherosclerosis Risk in Communities Study. *Stroke*. 2002;33:1487–1492.
4. Brownlee M, Aiello LP, Cooper ME, et al. Complications of diabetes mellitus. In: *Williams Textbook of Endocrinology*. Elsevier; 2016:1484–1581.
5. Early Treatment Diabetic Retinopathy Study Research Group. Treatment techniques and clinical guidelines for photocoagulation of diabetic macular edema: early treatment diabetic retinopathy study report Number 2. *Ophthalmology*. 1987;94:761–774.
6. Writing Committee for the Diabetic Retinopathy Clinical Research Network, Fong DS, Strauber SF, et al. Comparison of the modified Early Treatment Diabetic Retinopathy Study and mild macular grid laser photocoagulation strategies for diabetic macular edema. *Arch Ophthalmol*. 2007;125:469–480.
7. Hellstedt T, Vesti E, Immonen I. Identification of individual microaneurysms: A comparison between fluorescein angiograms and red-free and colour photographs. *Graefes Arch Clin Exp Ophthalmol*. 1996;234:S13–S17.
8. Niemeijer M, van Ginneken B, Cree MJ, et al. Retinopathy online challenge: automatic detection of microaneurysms in digital color fundus photographs. *IEEE Trans Med Imaging*. 2010;29:185–195.
9. Eftekhari N, Pourreza HR, Masoudi M, et al. Microaneurysm detection in fundus images using a two-step convolutional neural network. *Biomed Eng OnLine*. 2019;18:1–16.
10. Wu B, Zhu W, Shi F, et al. Automatic detection of microaneurysms in retinal fundus images. *Comput Med Imaging Graph*. 2017;55:106–112.
11. Long S, Chen J, Hu A, et al. Microaneurysms detection in color fundus images using machine learning based on directional local contrast. *Biomed Eng OnLine*. 2020;19:21.
12. Spencer T, Phillips RP, Sharp PF, Forrester JV. Automated detection and quantification of microaneurysms in fluorescein angiograms. *Graefes Arch Clin Exp Ophthalmol*. 1992;230:36–41.
13. Wang H, Chhablani J, Freeman WR, et al. Characterization of diabetic microaneurysms by simultaneous fluorescein angiography and spectral-domain optical coherence tomography. *Am J Ophthalmol*. 2012;153:861–867.e1.
14. Dubow M, Pinhas A, Shah N, et al. Classification of human retinal microaneurysms using adaptive optics scanning light ophthalmoscope fluorescein angiography. *Invest Ophthalmol Vis Sci*. 2014;55:1299–1309.
15. Horii T, Murakami T, Nishijima K, et al. Optical coherence tomographic characteristics of microaneurysms in diabetic retinopathy. *Am J Ophthalmol*. 2010;150:840–848.e1.
16. Schreur V, Domanian A, Liefers B, et al. Morphological and topographical appearance of microaneurysms on optical coherence tomography angiography. *Br J Ophthalmol*. 2019;103:630–635.
17. Kaizu Y, Nakao S, Wada I, et al. Microaneurysm imaging using multiple en face OCT angiography image averaging: morphology and visualization. *Ophthalmol Retina*. 2020;4:175–186.
18. Rand LI, Davis MD, Hubbard LD, et al. Color photography vs fluorescein angiography in the detection of diabetic retinopathy in the diabetes control and complications trial. *Arch Ophthalmol*. 1987;105:1344–1351.
19. Stitt AW, Gardiner TA, Archer DB. Histological and ultrastructural investigation of retinal microaneurysm development in diabetic patients. *Br J Ophthalmol*. 1995;79:362–367.
20. Frank RN. Diabetic retinopathy: current concepts of evaluation and treatment. *Clin Endocrinol Metab*. 1986;15:933–969.
21. Fukuda Y, Nakao S, Kaizu Y, et al. Morphology and fluorescein leakage in diabetic retinal microaneurysms: a

study using multiple en face OCT angiography image averaging. *Graefes Arch Clin Exp Ophthalmol*. 2022;260:3517−3523.
22. Lee SN, Chhablani J, Chan CK, et al. Characterization of microaneurysm closure after focal laser photocoagulation in diabetic macular edema. *Am J Ophthalmol*. 2013;155:905−912.e2.
23. Murakami T, Nishijima K, Sakamoto A, et al. Foveal cystoid spaces are associated with enlarged foveal avascular zone and microaneurysms in diabetic macular edema. *Ophthalmology*. 2011;118:359−367.
24. Ribeiro ML, Nunes SG, Cunha-Vaz JG. Microaneurysm turnover at the macula predicts risk of development of clinically significant macular edema in persons with mild nonproliferative diabetic retinopathy. *Diabetes Care*. 2013;36:1254−1259.
25. Early Treatment Diabetic Retinopathy Study Research Group. Grading diabetic retinopathy from stereoscopic color fundus photographs—an extension of the modified Airlie House classification. ETDRS report number 10. *Ophthalmology*. 1991;98:786−806.
26. Jia Y, Tan O, Tokayer J, et al. Split-spectrum amplitude-decorrelation angiography with optical coherence tomography. *Opt Express*. 2012;20:4710−4725.
27. Kraus MF, Potsaid B, Mayer MA, et al. Motion correction in optical coherence tomography volumes on a per A-scan basis using orthogonal scan patterns. *Biomed Opt Express*. 2012;3:1182−1199.
28. Wang J, Zhang M, Hwang TS, et al. Reflectance-based projection-resolved optical coherence tomography angiography. *Biomed Opt Express*. 2017;8:1536−1548.
29. Guo Y, Camino A, Zhang M, et al. Automated segmentation of retinal layer boundaries and capillary plexuses in wide-field optical coherence tomographic angiography. *Biomed Opt Express*. 2018;9:4429−4442.
30. Guo Y, Hormel TT, Xiong H, et al. Automated segmentation of retinal fluid volumes from structural and angiographic optical coherence tomography using deep learning. *Transl Vis Sci Technol*. 2020;9:54. https://doi.org/10.1167/tvst.9.2.54.
31. Parravano M, De Geronimo D, Scarinci F, et al. Diabetic microaneurysms internal reflectivity on spectral-domain optical coherence tomography and optical coherence tomography angiography detection. *Am J Ophthalmol*. 2017;179:90−96.
32. Parravano M, De Geronimo D, Scarinci F, et al. Progression of diabetic microaneurysms according to the internal reflectivity on structural optical coherence tomography and visibility on optical coherence tomography angiography. *Am J Ophthalmol*. 2019;198:8−16.
33. Khatri A, Pradhan E, Kc BK, et al. Detection, localization, and characterization of vision-threatening features of microaneurysms using optical coherence tomography angiography in diabetic maculopathy. *Eur J Ophthalmol*. 2021;31:1208−1215.
34. Chen R, Liang A, Yao J, et al. Fluorescein leakage and optical coherence tomography angiography features of microaneurysms in diabetic retinopathy. *J Diabetes Res*. 2022;2022:1−9.
35. Soliman W, Sander B, Hasler PW, Larsen M. Correlation between intraretinal changes in diabetic macular oedema seen in fluorescein angiography and optical coherence tomography. *Acta Ophthalmol*. 2008;86:34−39.
36. Moore J, Bagley S, Ireland G, et al. Three dimensional analysis of microaneurysms in the human diabetic retina. *J Anat*. 1999;194:89−100.